\documentstyle[12pt]{article}

\def\AdSs5{$AdS_5$}
\def\AdS5s5{$AdS_5 \times S^5$}
 
\def\al {\alpha^{\prime}}

\def\calZ{{\cal Z}}

\def\calN{{\cal N}} 
\def\Tr{\mbox{Tr}} 
\newcommand{\ie}{{\it i.e.~}}  
 
\newcommand{\s}{\sigma} 
 
\newcommand{\be}{\begin{equation}} 
\newcommand{\ee}{\end{equation}}  
\newcommand{\ba}{\begin{eqnarray}} 
\newcommand{\ea}{\end{eqnarray}} 
\newcommand{\ra}{\rangle}  
\newcommand{\la}{\langle} 
\newcommand{\ap}{\alpha^{\prime}}
\newbox\SlashedBox  
\def\fs#1{\setbox\SlashedBox=\hbox{#1} 
\hbox to 0pt{\hbox to 1\wd\SlashedBox{\hfil/\hfil}\hss}{#1}} 
\def\hboxtosizeof#1#2{\setbox\SlashedBox=\hbox{#1} 
\hbox to 1\wd\SlashedBox{#2}} 

\def\ms#1{\setbox\SlashedBox=\hbox{$#1$}
\hbox to 0pt{\hbox to 1\wd\SlashedBox{\hfil/\hfil}\hss}#1}

\def\R {\mbox{I}\!\mbox{R}}

\def\Z {\mbox{\sf{Z}}\hspace{-1.6mm}\mbox{\sf{Z}}\hspace{0.4mm}}
\begin{document}
     
\title{{\bf Yang--Mills Instantons vs. type IIB 
D-instantons\thanks{Talk presented by M.~Bianchi at the RFBR-INTAS School on
``Advances in Quantum Field Theory, Statistical Mechanics and Dynamical
Systems'', Landau Network at Centro A. Volta, Villa Olmo, Como, Italy,
September 7-20 1998, to appear in the proceedings.}}} 
\author{Massimo Bianchi and Stefano Kovacs \\
{\it Dipartimento di Fisica, \ \ Universit{\`a} di Roma \  
``Tor Vergata''} \\  {\it I.N.F.N.\ - \ Sezione di Roma \ ``Tor 
Vergata''} \\ {\it Via della Ricerca  Scientifica, 1} \\ 
{\it 00173 \ Roma, \ \ ITALY} }
\date{}	
\maketitle
\vspace{1cm}	 

\begin{abstract}
We report on the computation of the one-instanton contribution
to the 16-point Green function of fermionic composite operators
in ${\cal N}$=4 Super YM theory. The remarkable agreement, 
initially found in the case of an $SU(2)$ gauge group, 
with the one D-instanton contribution to the 
corresponding type IIB superstring amplitude on $AdS_5\times S^5$
is reviewed here. The recent extension by other authors of this result 
to any $SU(N)$ gauge group and to multi-instantons in the large $N$ limit 
is briefly discussed. We also argue that for the AdS/SCFT correspondence 
under consideration to work for any $N\geq 2$, at least for some protected 
interactions, string effects should require a truncation of the 
Kaluza--Klein spectrum for finite $S^5$ radius much in the same way as 
worldsheet unitarity restricts the allowed isospins in $SU(2)$ WZW models. 
Finally, we briefly comment on the logarithmic behavior of some four-point 
Green functions of scalar composite operators.
\end{abstract}	 

\vspace{3cm}

\underline{ROM2F/98/37}

\newpage

\section{Introduction} 
 
This talk will report on the correspondence between 
instanton effects in $\calN=4$ supersymmetric Yang--Mills theory in $D=4$
and D-instanton effects in type IIB superstring on anti de Sitter 
space. Most of the discussion here will rely on the  work \cite{bgkr}, 
for ease of presentation we will drop many numerical coefficients that 
have been carefully taken into account in that paper.  

Many of the recent non-perturbative 
insights in string as well as Yang--Mills theories are related to 
the properties of D-branes \cite{cjp}.
D-branes are BPS $p$-brane solutions of the 
type II (A or B) supergravity equations of motion 
that couple to antisymmetric tensors belonging to 
the Ramond-Ramond (R--R) sector. The massless R--R spectrum
of the type IIB superstring consists of a scalar, a two-form 
tensor and a self-dual four-form. These fields and their magnetic duals 
naturally couple to $p$-branes with odd $p$ ($p=-1,1,3,5,7,9$). 
In the rest of the talk we will be particularly interested in 
the type IIB D3- and D(-1)-branes. The latter are localized in space and 
(Euclidean) time and deserve the name of D-instantons \cite{greengut}.
A microscopic description of D$p$-branes is available
in terms of $p$-dimensional hypersurfaces on 
which open-strings can end \cite{cjp}. 
The $9-p$ transverse superstring coordinates 
satisfy Dirichlet boundary conditions whence the name.
Since the lowest lying modes of the open strings
comprise a vector supermultiplet, the world-volume dynamics of a $p$-brane 
is governed by the dimensional reduction of ${\cal N}$=1 SYM theory
from $D=10$ to $D=p+1$.
For a collection of $N$ coincident $p$-branes the world-volume gauge symmetry 
is enhanced to $U(N)$. The $U(1)$ factor describes the center-of-mass motion
and decouples from the rest \cite{wittdb}.

According to the proposal made by Maldacena in \cite{maldacena} 
$SU(N)$ $\calN=4$ supersymmetric Yang--Mills (SYM) theory in the 
large-$N$ limit is equivalent to type IIB supergravity on \AdS5s5. 
The SYM theory is located on the 
four-dimensional boundary of the five-dimensional anti de Sitter 
space and is interpreted as the 
world-volume theory for $N$ coincident D3-branes \cite{cjp}.  
The classical D3-brane soliton solution with $N$ units of
R--R 5-form charge is endowed with a constant dilaton, that 
determines the string coupling $g_s=e^{\phi_{_0}}$, and a metric
\be
ds^2 =  \left(1 + {L^4\over r^4} \right)^{-1/2} dx\cdot dx +  
\left( 1 + {L^4\over r^4} \right)^{{1}/{2}}   dy \cdot dy \quad ,
\label{d3brane}
\ee
where $(x^\mu, y^i)$  ($\mu=0,\dots 3$ and $i= 1,\dots,6$) 
are Cartesian coordinates (with $r^2 = y\cdot y$ the transverse distance) and
${L^4 = 4\pi g_s N {\al}^2}$.  
In the near-horizon limit $r\rightarrow 0$ or, better, for
$L^{4} >> {\ap}^{2}$ (\ref{d3brane}) tends to the metric on 
\AdS5s5\ described by 
\be  
ds^2 =  {L^2 \over \rho^2}(dx\cdot dx  + d\rho^2) + d{\omega_5}^2  ,
\label{cfmetric} 
\ee 
where  $\rho^2 =L^4/r^2$ and $d\omega_5^2$ 
is the spherically-symmetric  constant curvature metric on $S^5$.

The $SU(4)\times SO(4,2)$ isometry of the type IIB background (\ref{cfmetric})
is the bosonic part of the exact superconformal symmetry of the $\calN=4$ SYM 
theory. The (conjectured) exact $SL(2,\Z)$ ``S-duality" 
of the type IIB theory corresponds to the (conjectured) exact electromagnetic 
duality \cite{olivemont} of four-dimensional ${\cal N}$=4 
SYM theory \cite{bssgos}. The former is connected to the 
existence of an infinite set of stable dyonic BPS string solitons 
\cite{schwarz}, the latter
to the existence an infinite set of stable dyonic BPS states \cite{sen}. 
The abelian Coulomb phase of the SYM theory corresponds to separating the 
D3-branes or, equivalently, to giving non-vanishing vacuum expectations 
values (vev's) to the scalar fields in the Cartan subalgebra.
The resulting theory is free and the superconformal invariance is broken.
The phase in which all scalar fields 
have vanishing  vev's describes a highly nontrivial 
superconformal field theory \cite{gubskleb,maldacena,gkp,wittone}.
Superconformal invariance is protected against quantum corrections by
the exact vanishing of the renormalization group $\beta$-function \cite{fz}.    
The complexified coupling constant of the $\calN=4$ SYM 
theory is related to the one of the type IIB superstring on \AdS5s5\ by 
\begin{equation} 
\label{dict} g_s = {g_{_{YM}}^2 \over {4\pi}} \:, \qquad
\chi_{_{0}} =  {\theta_{_{YM}} \over 2\pi} \:,   
\end{equation}   
where $\chi_{_{0}}$ is the constant  R--R axionic background,
$g_{_{YM}}$ is the SYM coupling and $\theta_{_{YM}}$ is the 
SYM vacuum angle. 
 
Long time before any D-brane argument was put forward, the
maximally supersymmetric compactification of the type IIB superstring on the
\AdS5s5\ background was considered by Schwarz \cite{schw}
and the spectrum of excitations was studied in great detail 
\cite{kiromvan}.
It is thus natural to make Maldacena's proposal more precise and suggest that
the boundary values of the fields of the bulk type IIB superstring 
are sources that couple to gauge-invariant composite operators of the 
four-dimensional boundary $\calN=4$ superconformal field theory (SCFT)
\cite{gkp,wittone,adf,ff}. The 
lowest lying states, assembled into the ${\cal N}$=8 gauged supergravity 
multiplet, couple to the supermultiplet of SYM currents
that are ``bilinear'' in the elementary SYM fields. 
Higher $\ell$ spherical harmonics, \ie Kaluza--Klein (KK) descendants
of the supergravity fields, couple to other chiral primary fields,
that correspond to ``order $\ell$'' gauge-invariant polynomials  
in the elementary SYM fields. For finite $N$, \ie for finite radius, 
one expects a truncation to $\ell \leq N$ that should result by taking
into account full-fledged stringy geometry as in any string compactifications. 
For instance, in the compactification on the $SU(2)$ group manifold
unitarity of the world-sheet theory restricts the allowed isospins
of the primary fields to the range $j=0,\dots k/2$. 
The level $k$ of the current algebra is related to the radius of $S^3$.
Similarly the number $N$ of D3-branes is related to the radius of $S^5$.
In the case at hand, $AdS_5\times S^5$, despite some interesting 
progress \cite{gsads}, a useful world-sheet description is still lacking.
Nevertheless the truncation of the KK spectrum seems to be a robust result
that matches with the naive expectation on the SYM side.
Moreover, stable string excitations that couple to non-chiral primary fields
are needed for the closure of the operator algebra \cite{konishi}.

In this picture, the effective action of the type IIB superstring is 
the generating  functional of connected  
gauge-invariant correlation functions in the SYM theory.  
The semiclassical supergravity approximation 
is valid at length scales much larger than the string 
scale  or, equivalently, in the limit $g_{s} N >> 1$.  
From the SYM viewpoint, this coincides with a subcase of the large-$N$ 
limit introduced by 't Hooft \cite{thooft} in which the coupling  $\hat g^2 =   
g_{_{YM}}^2 N$ is fixed at a large value. By taking into account string 
corrections to the low energy supergravity approximation,
the AdS/SCFT correspondence under consideration
is expected to work for finite values of $N$ as well.
Indeed a surprising agreement between SYM instanton computations for
$N=2$ and some D-instanton corrections to the type IIB superstring
effective action has been shown in \cite{bgkr}.

The explicit connection between the bulk theory and the boundary  
theory can be  expressed symbolically as \cite{wittone,gkp,ffz}
\be 
\int [D\Phi]_J\exp(-S_{_{IIB}} [\Phi]) = \int [DA] \, \exp(-S_{_{YM}}[A] + 
{\cal O}[A] J) \,  , 
\label{main} 
\ee 
where $S_{_{IIB}}$ is the effective action of the type IIB superstring
which is evaluated  in terms of the `massless' supergravity fields, 
their Kaluza--Klein descendents, and possibly the string excitations.
We have collectively denoted the bulk fields by 
$\Phi(z;\omega)$, where $\omega$ are the coordinates on 
$S^5$ and $z^M\equiv (x^\mu,\rho)$ ($M=0,1,2,3,5$ and $\mu=0,1,2,3$) the 
coordinates on $AdS_5$ ($\rho \equiv z_5$ is  
the coordinate transverse to the boundary).  The functional integral 
depends on the boundary values, $J(x)$, of the bulk fields.  
Elementary fields of the $\calN=4$ SYM theory on the boundary   
are denoted by $A$ and ${\cal O}(A)$ are 
gauge-invariant  composite operators to which $J(x)$'s 
couple. The conformal dimension $\Delta$
of an operator 
${\cal O}$ is related to the AdS `mass' $m$ of the corresponding 
bulk field $\Phi$. For scalars one has $(mL)^2 = \Delta(\Delta -4)$.  
Only the positive branch of $\Delta_{\pm} = 2 \pm \sqrt{4+(mL)^2}$ 
is relevant for the lowest-`mass' supergravity multiplet. 

The correct recipe for computing correlation functions amounts
to computing the truncated Green functions in the bulk and attaching
bulk-to-boundary propagators to the external legs \cite{gkp,wittone,freedman}. 
The precise forms of the latter
depend on the spin and `mass' of the field. For instance, 
the bulk-to-boundary Green function for a bulk scalar field 
$\Phi_{m}$ associated to a dimension $\Delta$ operator ${\cal O}_{\Delta}$ 
on the boundary reads
\be
 K_{\Delta} (x^\mu, \rho; x'^{\mu},0) =  
{\rho^{\Delta} \over  (\rho^2 
+ (x-x')^2)^{\Delta}} \quad , 
\label{greenfun} 
\ee
up to an overall normalization constant.
For simplicity we have suppressed the $\omega$ dependence so that 
the expression (\ref{greenfun}) is appropriate for 
an `S-wave' process in which there are no excitations 
in the directions of the five-sphere, $S^5$.  
In terms of $K_{\Delta}$ the bulk field 
\be
\Phi_m (z;J) = \int d^4x' K_{\Delta} (x,\rho; x',0) J_{\Delta}(x') 
\label{btob}
\ee
satisfies the boundary condition $\Phi_m (x,\rho;J) \approx 
\rho^{4-\Delta} J_{\Delta}(x)$ as $\rho \to 0$,
since $\rho^{\Delta-4}K_{\Delta}$ reduces to a $\delta$-function on the 
boundary.   
 
For our considerations it will prove crucial that the
expression  (\ref{greenfun}) in the case $\Delta =4$ has exactly 
the same form as the 
contribution of a Yang--Mills (YM) instanton to $\Tr  (F^-_{\mu\nu})^2$ 
(where $F^-_{\mu\nu}$ is the non-abelian self-dual field strength) when 
the  fifth  
coordinate $\rho$ is identified with the instanton scale.  At the 
same time, it has been shown in \cite{bgkr}
that in this case (\ref{greenfun}) has  precisely the 
same form as the profile 
of a five-dimensional D-instanton centered at the point $z^M$  in  
\AdSs5\  and evaluated at the boundary point $(x^{\prime \mu},0)$.   
This is a key observation in  identifying  D-instanton  effects of 
the bulk theory with YM instanton effects in  the boundary 
theory.  It is related to the fact that the moduli space of a
YM instanton contains an $AdS_5$ factor.
  
Two-point and three-point correlation functions of superconformal 
currents are not renormalized from their  free-field  values  due to  the 
$\calN=4$ superconformal invariance so   
they  do not get interesting interaction corrections 
\cite{freedman,afgj,howest,sei}. However,  
higher-point correlation functions do receive nontrivial interaction 
corrections.  In this talk we will be concerned with calculations of 
processes in which  
the one-instanton contributions can be evaluated exactly  to lowest 
order in perturbation theory. 
The contributions of a one-instanton configuration to some correlation 
functions in the $SU(2)$ SYM theory have been evaluated in \cite{bgkr}. 
The result for the 16-point function, 
$\langle {\widehat \Lambda}(x_1) \dots {\widehat \Lambda}(x_{16})\rangle$, 
of gauge-invariant fermionic composite operators ${\widehat \Lambda}=
F^-\lambda$, where $F^-$ is the self-dual field strength 
and $\lambda$ is the spin-1/2 gaugino, has been later generalized to any $N$
\cite{dm}. More importantly, the $K$-instanton contributions have been 
evaluated in the large $N$ limit by means of a saddle-point approximation of 
the integral over the relevant moduli space \cite{mattis2}.

The common feature of all the
correlation functions considered in \cite{bgkr,dm,mattis2}, 
as well as many others  that  are related to them by supersymmetry, 
is that they provide precisely the sixteen `geometric'
fermionic zero modes that are needed to give a nonzero result in the 
instanton background.   

In \cite{bgkr} it has been shown that the form of the   
leading charge-one  D-instanton contribution to 
 the $\Lambda^{16}$  amplitude in the type IIB superstring  
matches with  the expression for the sixteen-${\widehat \Lambda}$
correlator calculated in the $\calN=4$ $SU(2)$ SYM theory.
Given this agreement, that have been further confirmed in \cite{dm}
and definitively established in \cite{mattis2}, 
supersymmetry guarantees that all 
of the related correlators must also agree with their AdS counterparts.  
Although 
a detailed comparison for all such correlation functions has not been
completed, the correspondence has been outlined in \cite{bgkr,mattis2}.

In order to further clarify the correspondence between YM
instantons and D-instantons, the classical D-instanton 
solution of the type IIB supergravity equations in the 
\AdS5s5\ background has been explicitly determined in \cite{bgkr}. 
The solution is particularly simple to establish by 
using the fact that \AdS5s5\ is conformally flat, which relates the 
solution to the flat-space solution of \cite{ggp}. 
Once the D-instanton solution in \AdS5s5\ is known one can directly compute the
semi-classical approximation to the $\Lambda^{16}$ correlation function
and check the agreement with the previous computations.

\section{${\cal N}$=4 SYM in $D$=4} 
 
The field content of the maximally supersymmetric (${\cal N}$=4)  
four-dimensional Yang--Mills \cite{bssgos,fz,foursusy} theory is unique 
apart from the choice  of the gauge group. As mentioned above,  
the theory is  classically invariant under superconformal 
transformations as well as  under global $SU(4)$ transformations, 
which form the R-symmetry 
group of  internal automorphisms of the ${\cal N}$=4 superconformal algebra.  
There is also an external $U(1)_B$ automorphism under which 
the Poincar\'e and conformal left supercharges transform with opposite 
charge \cite{int}. 
The field content  consists of six real scalars,  four Weyl 
spinors and one vector which are all in the adjoint  representation 
of  the gauge group. More precisely, the scalars $\varphi^{AB} = - 
\varphi^{BA}$ (with 
${\overline 
\varphi}_{AB}=\frac{1}{2}\varepsilon_{ABCD}\varphi^{CD}$)  are in 
the {\bf 6} of the 
$SU(4)$ R-symmetry group\footnote{The superscripts $A,B=1,\dots,4$ 
label {\bf 
4}'s of $SU(4)$ while  subscripts label the {\bf 4}$^*$'s.}.  In the 
following we will also use  
the notation $\varphi^{i}=\frac{1}{2} {\overline{t}^{i}}_{AB} 
\varphi^{AB}$ ($i=1,\ldots,6$) where  $\varphi^{i} =  
{\varphi^{i}}^{*}$ and $\bar t^i_{\ AB}$ are  
Clebsch--Gordan coefficients that  couple two 
{\bf 4}'s to a {\bf 6} (these are six-dimensional generalizations of 
the four-dimensional
${\sigma^{\mu}}_{\alpha {\dot \alpha}}$ matrices).   The spinors  
$\lambda^{A},~{\overline \lambda}_{B}$ transform  as {\bf 4}  and 
{\bf 4}$^{*}$, respectively, and the vector $A_{\mu}$ is a singlet 
of $SU(4)$. All couplings are proportional to $g_{_{YM}}$ and to 
$f_{abc}$.

The extended supersymmetry transformations with parameters 
${\eta_{\alpha}}^{A}$ and ${\overline\eta^{\dot\alpha}}_{A}$ are given 
in component-field notation by
\begin{eqnarray} 
\delta \varphi^{i} &=& \frac{1}{2} \left(
\bar{t}^i{}_{AB}{\lambda^{\alpha}}^{A} {\eta_{\alpha}}^{B} 
+t^{iAB}{{\overline{\eta}}_{\dot\alpha}}_{A}{\overline\lambda}^{\dot\alpha}_{B} 
\right) \nonumber \\ 
\delta {\lambda_{\alpha}}^{A} &=& -\frac{1}{2} F^-_{\mu \nu}  
{{\sigma^{\mu \nu}}_{\alpha}}^{\beta} {\eta_{\beta}}^{B} + 
2i (\ms{D}_{\alpha {\dot \alpha}} \varphi^{i})
t_i{}^{AB} {\overline{\eta}}^{{\dot \alpha}}_{B} + 
2g_{_{YM}}[{{\varphi}}^{i}, \varphi^{j}] t_{ij}{}^A{}_B {\eta_{\alpha}}^{B}  
\nonumber \\ 
\delta A_{\mu} &=& -i {\lambda^{\alpha}}^{A}{\sigma^{\mu}}_{\alpha{\dot\alpha}} 
{{\overline \eta}^{\dot \alpha}}_{A} 
-i {\eta^{\alpha}}^{A}{\sigma^{\mu}}_{\alpha {\dot \alpha}} 
{{\overline \lambda}^{\dot \alpha}}_{A}  \quad , 
\label{transs} 
\end{eqnarray} 
where as usual $t_{i_1\cdots i_p}$ denote $p$-fold 
antisymmetric products of $t_i$ (and $\bar{t}_j$) matrices with strength-one. 
The classical theory is superconformally invariant 
\cite{foursusy} and  this 
property is believed to be preserved at the quantum level thanks to 
the 
exact vanishing of the $\beta$-function \cite{fz}. The Noether  currents 
associated with the superconformal transformations,  together with  those 
corresponding to 
chiral $SU(4)$ transformations, constitute a supermultiplet that
is bilinear in the elementary fields in the abelian case. The 
current supermultiplet includes the  energy-momentum tensor, 
${\cal T}_{\mu\nu}$, the 
supersymmetry  currents, ${\Sigma^{\mu}}_{\alpha A}$, and  
the $SU(4)$ R-symmetry currents,
${{{\cal J}^{\mu}}_{A}}^{B}$. 
The remaining components are obtained by on-shell 
supersymmetry transformations. They 
consist of three  scalar components  (${\cal C}$, ${\cal E}^{(AB)}$, 
${\cal Q}^{ij}$), two fermionic spin-1/2 components  
(${{\widehat \chi}^C}_{AB}$ and ${\widehat \Lambda^A}$) 
and  one antisymmetric tensor (${\cal B}_{\mu\nu}^{[AB]}$). 
In the non-abelian case, aside from the obvious trace over the colour 
indices that enters, for instance, the definition of
\be
\label{qdef}
{\cal Q}^{ij}=\Tr 
\left(\varphi^i\varphi^j-{1\over 6} \delta^{ij} \varphi_k\varphi^k \right)
\quad ,
\ee
there are additional terms that are not always exhausted, as in
the definition
\be
{{\widehat \Lambda}_{\alpha}}^{A} = {1\over 2}
\Tr \left( {F^{-}}_{\mu \nu} {{\sigma^{\mu \nu}}_{\alpha}}^{\beta}
{\lambda_{\beta}}^{A} \right) \quad ,
\label{lambdadef}
\ee
by the covariantization of all the derivatives. In particular, 
the complete non-abelian expression for ${\cal E}^{(AB)}$ \cite{bgkr}, 
\be 
{\cal E}^{(AB)} = 
\Tr \left( \lambda^{\alpha A} {\lambda_{\alpha}}^{B} \right) + 
g_{_{YM}} \, t_{[ijk]}^{(AB)_+} 
\Tr \left( \varphi^{i}\varphi^{j}\varphi^{k} \right) \quad , 
\label{exacte} 
\ee 
has played a crucial r\^ole in showing a one-loop 
non-renormalization theorem for the three point functions of 
descendants of the chiral primary fields ${\cal Q}$ \cite{dhfs}.

It is often useful to decompose the ${\cal N}$=4 multiplet in  terms 
of either ${\cal N}$=1 or ${\cal N}$=2 multiplets.  
We will need the ${\cal N}$=2 description in the following. 
The global symmetry that is manifest  in this description is 
$SU(2)_{{\cal V}}\times SU(2)_{{\cal H}} \times U(1)$ and the ${\cal 
N}$=4 elementary supermultiplet   
decomposes into a $\calN=2$ vector multiplet ${\cal V}$ and a 
hypermultiplet,  
${\cal H}$.  
We will denote  the representations of $SU(2)_{\cal V} \times 
SU(2)_{\cal H} 
\times U(1)$  by  ({\bf r}$_{{\cal V}}$, {\bf r}$_{{\cal 
H}}$)$_{q}$  with the subscript $q$ referring to  the  
$U(1)$ charge  and will make use of the following notation for the  
component 
fields, 
 \begin{eqnarray} 
 	{\cal V} ~~~ && \rightarrow ~~~ \lambda^{u} \in {({\bf 2}, 
        {\bf  1})}_{+1}, 
 	~~~\varphi \in {({\bf 1}, {\bf 1})}_{+2}, ~~~A_\mu \in {({\bf 1}, 
        {\bf 1})}_{0} \nonumber \\ 
 	{\cal H} ~~~ && \rightarrow ~~~ \psi^{{\dot u}} \in {({\bf 1},  
 	{\bf 2})}_{-1}, ~~~ q_{u{\dot u}} \in {({\bf 2}, {\bf 2})}_{0} ~~,  
 	\label{entwo}
 \end{eqnarray}  where $u,{\dot u} = 1,2$. 

In calculating the correlation functions it is 
important to understand the systematics of the fermion zero modes  
associated with the YM instanton.  These can  either  be 
determined directly from the supersymmetry  transformations 
(\ref{transs}) or by making use of the  explicit  Yukawa couplings 
given in \cite{bgkr}. In the following we will 
review the computation of correlation functions of currents in 
the $\calN=4$ SYM theory which receive contributions 
in the one-instanton semi-classical approximation. 
As in \cite{bgkr} we will mainly restrict our discussion  
to the case of an $SU(2)$ gauge group although we will make some qualitative
comments about the $SU(N)$ case following \cite{dm} and briefly mention
the steps required for the computation of $K$-instanton contributions
in the large $N$ limit following \cite{mattis2}.
The special feature shared  by the particular  
observables that we will consider is that they 
can saturate the 16 supersymmetric plus superconformal 
gaugino zero modes that arise in the instanton background.

\section{Type IIB superstring in $D$=10} 
  
We now turn to consider the other side of the correspondence, \ie
the type IIB superstring. In $D=10$, 
this theory enjoys chiral ${\cal N}$=(2,0)
supersymmetry associated to the presence of a complex gravitino 
of positive chirality, that
transforms with charge $+1/2$ under an anomalous $U(1)_B$ R-symmetry. 
The massless bosonic spectrum includes the metric, 
a scalar dilaton, $\phi$, and 
a two-form potential in the Neveu-Schwarz -- Neveu-Schwarz (NS--NS) 
sector and a pseudo-scalar, $\chi$,
another antisymmetric tensor and a four-form potential
whose field-strength is constrained to be self-dual, 
in the R--R sector. The classical theory
is invariant under $SL(2,\R)$ transformations that act projectively on
the complex scalar field $\tau = \chi+ie^{-\phi}$. The two 
antisymmetric tensors form a doublet, while the metric (in Einstein frame!)
and the self-dual four-form potential are inert.
At the quantum level the classical continuous symmetry is expected to be 
broken to $SL(2,\Z)$. Due to the self-duality constraint finding the full 
action including the four-form potential is far from obvious \cite{pst}.
For our purposes we only need to display the two leading terms in the 
derivative expansion of the type 
IIB effective action that involve the (derivatives 
of the) metric and some overall function of the complex scalar $\tau$.
In string frame they are given by \cite{greengut}, 
\ba
&&(\alpha')^{-4} \int d^{10}X \sqrt g \left(e^{-2\phi} R + \kappa 
(\alpha^\prime)^3 f_4(\tau,\bar\tau) 
e^{-\phi/2} {\cal R}^4    \right) \nonumber\\
&& =  L^{-8} \int d^{10}X \sqrt g \left((4\pi N)^2 R + \kappa L^6 (4\pi 
N)^{1/2}
 f_4(\tau,\bar\tau)  {\cal R}^4    \right)  \: ,
\label{twoterms}
\ea
where $\kappa$ is a known constant and the relation 
$L^4 = 4\pi g_s {\ap}^2$ has been substituted in the second line.

The Riemann curvature enters the ${\cal R}^4$ term in a manner 
that only involves the Weyl tensor. In \cite{banksgreen} it was pointed
out that the ${\cal R}^4$  term vanishes in the \AdS5s5\ background 
because it involves a fourth power of the (vanishing) Weyl tensor.
The first, second and third functional derivatives of ${\cal R}^4$
vanish as well, and as a result one finds no corrections to zero, one,
two and three-point amplitudes.
However, there is a non-zero four-graviton amplitude  
arising from this term. The boundary values of these gravitons are
sources for various bosonic components of the SYM current 
supermultiplet.  For example, the components of the metric  in the
$AdS_5$ directions couple to the stress tensor, ${\cal T}^{\mu\nu}$,
whereas the traceless components polarized in the $S^5$ directions couple to  
massive Kaluza-Klein states. A linear combination of the ``quadrupole moments"
of the trace of the
metric on $S^5$ and the fluctuations of the R--R four-form potential
couples to the scalar composite operator, ${\cal Q}^{ij}$.  
 
We will be particularly concerned with a sixteen-fermion 
interaction \cite{greengutkwon}, that is related to 
 ${\cal R}^4$ by supersymmetry. It appears at the same (subleading)
order in the derivative expansion and reads
\be
\label{lambint} 
(\alpha')^{-1} \int d^{10}X \sqrt g \, e^{-\phi/2} 
f_{16}(\tau,\overline{\tau})  \Lambda^{16}   + \mbox{c.c.} \, ,
\ee
 where $\Lambda$ is a complex spin 1/2 dilatino of negative chirality 
 which transforms under the $U(1)_B$ R-symmetry with charge +3/2 and the 
interaction is antisymmetric in the sixteen spinor indices. 
Supersymmetry arguments \cite{pk,grse} imply that the functions $f_{16}$ and 
$f_4$ are related by the action of $SL(2,\Z)$ modular covariant derivatives.
Whereas $f_4$ transforms with modular weight $(0,0)$, the function $f_{16}$ has 
weight $(12,-12)$  and therefore transforms with a non-trivial
phase under $SL(2,\Z)$.   This is precisely the phase required to
compensate for the anomalous $U(1)_B$ transformation of the 16 $\Lambda$'s 
so that the full expression (\ref{lambint}) is invariant under 
the $SL(2,\Z)$ S-duality group. In the small coupling limit, 
all of these terms can be expanded in the form 
\be 
e^{-\phi/2}  f_n = a_n \zeta(3)e^{-2\phi}  +  
b_n + \sum_{K =1}^\infty \mu_{K} (K e^{-\phi})^{n -7/2} e^{2\pi i K\tau} 
\left(1 + \sum_{k=1}^\infty c_{k,n}^K (K e^{-\phi})^{-k} \right) \; . 
\label{fnexp} 
\ee  
Whereas the first two terms have the form of string tree-level 
 and one-loop contributions, the last one 
has the appropriate weight of $e^{2\pi i K\tau}$ 
to be associated with a charge-$K$ D-instanton effect. 
Anti-D-instanton contributions have not been displayed.
The coefficients $c_{k,n}^K$ as well as the `D-instanton partition functions'
$\mu_{K}$ are explicitly given in \cite{greengutkwon}. 
The coefficient of the leading term in the series of perturbative  
fluctuations around a charge-$K$ D-instanton is independent of which 
particular interaction term is being discussed and reads 
\be 
\calZ_K = \mu_{K} (K e^{-\phi})^{-7/2} e^{2\pi i K\tau} \: .
\label{znew} 
\ee  
It should be identified with the contribution of a charge-$K$  
D-instanton to the measure in string frame which,  
up to an overall numerical factor, reads
\be
\label{dmeasdef}
d\Omega_K^{(s)} = (\alpha')^{-1} d^{10} X \, d^{16}\,  \Theta \,{\cal Z}_K \, .
\ee
In doing this we are being cavalier about the fact that the full series 
is not convergent (it is actually an asymptotic
approximation to a Bessel function).  In the end, consistency of the 
full theory, and in particular modular invariance, should require 
considering the complete expression for the instanton contribution. 
  
As observed in \cite{banksgreen}\ the charge-$K$  
D-instanton action that appears in the exponent in (\ref{znew}) 
coincides with the action of a charge-$K$ YM instanton in 
the boundary theory. This indicates a correspondence between 
these sources of non-perturbative 
effects, that is reinforced by the correspondence 
between other factors.  For example, after substituting 
$e^\phi = g_{_{YM}}^2/4\pi$ and 
$\alpha' = g^{-1}_{_{YM}} L^2 N^{-1/2}$, the measure 
(\ref{dmeasdef}) contains an overall factor of the coupling constant 
in the form $g_{_{YM}}^8$.
Indeed, this is exactly the power expected on the
 basis of the  \AdS5s5/SYM\  correspondence   since
the one-instanton contribution to the Green functions
in the ${\cal N}$=4 SYM theory also has a factor of 
$g_{_{YM}}^8$ arising from the combination  of the bosonic and 
fermionic zero modes norms.   
Following \cite{bgkr} we will pursue this issue further in the next section 
by comparing  the leading 
instanton contributions  to  type IIB superstring amplitudes with the  
corresponding  $\calN=4$ current correlators.

\section{Testing the AdS/SCFT correspondence} 
 
The initial tests of Maldacena's conjecture 
have amounted to the computation of two-point and three-point 
correlations of primary fields \cite{twoandthree}.
Up to an overall constant these are fixed by superconformal invariance
and the tests amount to checking the consistency of the relative normalizations
and the (non)-vanishing of some three-point couplings.
Four-point functions, on the contrary, display some interesting dynamical
issues that we will briefly address in the last part of the talk.

Momentarily, we review the computation of the one-instanton contribution to the 
SYM correlation function $G_{16} = \la 
\widehat\Lambda(x_{1})\ldots\widehat\Lambda(x_{16})\ra$ 
for an $SU(2)$ gauge group 
and its comparison with the correspondent D-instanton contribution to the 
$\Lambda^{16}$ amplitude in the type IIB superstring. In 
the case of the bulk type IIB theory, D-instanton
effects may be either extracted directly from 
the exactly known terms in the effective action or deduced
from the integration over the semi-classical fluctuations 
around the $AdS_5\times S^5$ D-instanton solution.  

\subsection{Super Yang--Mills description}

The contribution of the YM one-instanton sector to the correlation function 
\begin{equation} 
G_{16}(x_p) = \langle \prod_{p=1}^{16} g_{_{YM}}^2 
\widehat \Lambda^{A_{p}}_{\alpha_{p}}(x_{p})  
	\rangle_{_{K=1}} \; , 
	\label{l16}
\end{equation} 
of sixteen fermionic superconformal currents 
${\widehat \Lambda}_\alpha^{A}$, defined in (\ref{lambdadef}),
is particularly simple to analyze. 
Since each $\widehat\Lambda$ insertion
provides a single fermion zero mode it is necessary to consider the 
product of sixteen $\widehat\Lambda$'s in order  to saturate the sixteen  
grassmannian integrals. To leading order in $g_{_{YM}}$, 
$G_{16}$ does not receive contribution from anti-instantons.
The leading term is simply obtained by replacing  each $F^{-}_{\mu \nu}$ 
with the $SU(2)$ instanton solution  
\be 
 F^{a-}_{(0)\mu\nu} =  - \frac{4}{g_{_{YM}}} 
{\eta_{\mu\nu}^a \rho_0^2 \over \left(\rho_0^2  
+ (x-x_{0})^2 \right)^2}  \: , 
\label{instform}
\ee  
where $\eta^a_{\mu\nu}$ is the 't Hooft symbol \cite{thooftb},  
and each $\lambda_\alpha^{\ A}$ with the corresponding 
zero mode, ${\lambda_{(0)\alpha}}^{A}$. The latter 
can be deduced from the action of the supersymmetry transformations
that are broken in the instanton background. The relevant transformation
may be found in the second line of (\ref{transs}) and gives
\be
\label{lamzero}
\lambda_{(0) \alpha}^A =  {1\over 2} 
F_{(0)\mu\nu}^- \sigma_\alpha^{\mu\nu\, \beta} 
{1 \over \sqrt{\rho_0}} \left( \rho_0\eta_\beta^A + 
(x-x_0)_\mu \sigma^\mu_{\beta\dot \beta }
\bar\xi^{\dot \beta A} \right) \; .
\ee
Up to an overall numerical coefficient, see \cite{bgkr},  
the resulting correlation function has thus the form 
\begin{eqnarray} 
  G_{16}(x_p) 
  &=& g_{_{YM}}^{8} e^{-{8\pi^2 \over g_{_{YM}}^2} + i\theta_{_{YM}}} 
  \int \frac{d^{4}x_{0} \, d\rho_0}{\rho_0^{5}}  
  \int d^{8}\eta d^{8}{\overline \xi}  \nonumber \\  
  && \prod_{p=1}^{16} \left[ 
  \frac{\rho_{0}^{4}}{[\rho_0^{2}+(x_{p}-x_{0})^{2}]^{4}} \,
  \frac{1}{\sqrt{\rho_{0}}}
  \left( \rho_0 \eta^{A_{p}}_{\alpha_{p}}+
  {(x_{p}-x_{0})}_{\mu}  \sigma^{\mu}_{\alpha_{p}{\dot \alpha}_{p}}  
  {\overline \xi}^{{\dot \alpha}_{p}A_{p}} \right)  \right] \: . 
\label{l16-sym} 
\end{eqnarray} 
The integration over the fermionic collective coordinates leads to an 
$SU(4)$ and Lorentz invariant sixteen-index tensor, $t_{16}$. 
We will leave the 
expression in the unintegrated form (\ref{l16-sym})  which will be 
momentarily compared with the corresponding expression obtained in type
IIB superstring theory on \AdS5s5.  Again,  the resemblance of the instanton
form factor to the AdS Green function will be of significance here.

The above computation has been recently extended to generic $N$ 
\cite{dm} and to higher instanton number in the large $N$ limit \cite{mattis2}.
We can not help briefly reviewing the main steps of these extensions here.
Since for a gauge group $SU(N)$ there are $2N\cdot{\cal N}$ gaugino zero modes 
in the one-instanton background it might appear from a superficial analysis 
that the $\widehat \Lambda^{16}$-correlation function should vanish for 
$N\ge2$. However it has been argued in \cite{bgkr} and then clearly shown 
in \cite{dm} that only the sixteen 
`geometric' (8 supersymmetric + 8 superconformal)
zero-modes are actually relevant in the general case of $SU(N)$, 
since all the remaining ones, $\mu_{i}^{A}$ are `lifted' by a four-fermion 
term that is generated in the one-instanton action. The instanton action 
computed in \cite{dm} has the form
\be
S_{_{\mbox{\tiny{inst}}}}=\frac{8 \pi^2 K}{g_{_{YM}}^2} + 
S_{_{\mbox{\tiny{4F}}}} \, ,
\label{liftaction}
\ee
where $S_{_{\mbox{\tiny{4F}}}}$ contains a quartic fermionic interaction 
built up with the $8N-2$ additional fermionic collective coordinates. 
The exact form of $S_{_{\mbox{\tiny{4F}}}}$ in the $K=1$ sector is
\be
S_{_{\mbox{\tiny{4F}}}}=\frac{\pi^2}{16\rho^2 g_{_{YM}}^2} \, 
\varepsilon_{ABCD} \left[
\sum_{i=1}^{N-2}{\overline \mu}^A_i \mu^B_i\right] \left[
\sum_{j=1}^{N-2}{\overline \mu}^C_j \mu^D_j\right] \, .
\label{4faction}
\ee
$S_{_{\mbox{\tiny{4F}}}}$ is supersymmetric and lifts all but the 16 
`geometric' gaugino zero modes. In computing the functional integral
for the Green function it is convenient to perform a Hubbard-Stratonovich
transformation of the fermion bilinears and represent 
$S_{_{\mbox{\tiny{4F}}}}$ as a
gaussian integral over auxiliary bosonic `collective coordinates' $\chi^i_{AB}$
\cite{dm}. As a result the expression for the 
$\widehat \Lambda^{16}$-correlator reviewed above for the case $N=2$ 
is actually true for all $N$, 
including an overall $N$-dependent factor that takes into 
account the scaling with $\ap$ in the type IIB description.
The auxiliary bosonic `collective coordinates' $\chi^i_{AB}$ play a crucial
role in the evaluation of $K$-instanton contributions. 
In \cite{mattis2} it is shown that after a saddle-point approximation
the $\chi$'s
can be viewed as coordinates on $K$ copies of $S^5$ thus completing
the identification of the YM-instanton moduli space (for gauge-invariant 
observables) in the large $N$ limit with the type IIB D-instanton moduli space!

\subsection{Type IIB description}

The first method for obtaining the correspondent $\Lambda^{16}$-amplitude 
in the type IIB description is to 
expand the function $f_{16}$ in (\ref{lambint}) to extract the
one-instanton term.  In order to compare with the SYM 
sixteen-point correlator we need to consider the situation in which
all sixteen fermions propagate to prescribed configurations
on the boundary.  
The Dirac operator acting on spin-1/2 fields in AdS$_{5}$ was given 
in \cite{henningsfet} by  
\begin{equation} 
  \ms{D}\Lambda  = {e_{{\hat L}}}^{M} \gamma^{\hat{L}}  
  \left( \partial_{M} +\frac{1}{4} 
  \omega^{\hat{M}\hat{N}}_{M} \gamma_{\hat{M}\hat{N}}
  \right)\Lambda
  =(\rho  \gamma^{\hat 5}  
  \partial_{5}  
  + \rho \gamma^{\hat\mu} \partial_{\mu} - 2 \gamma^{\hat 5})
  \Lambda \; , 
\label{dirac-op} 
\end{equation} 
where ${e_{\hat{L}}}^{M}$ is the vielbein and 
$\omega^{\hat{M}\hat{N}}_{M}$  the  
spin connection for the AdS part of the metric (\ref{cfmetric}) 
(hatted indices refer to the tangent space), 
and $\gamma^{ \hat\mu}$ are the  four-dimensional Dirac matrices.  
Equation (\ref{dirac-op}) leads to the normalized 
bulk-to-boundary propagator of the fermionic field $\Lambda$ 
of AdS `mass' $m=-3/2L$, associated to the composite operator  
${\widehat \Lambda}$ of dimension $\Delta = \frac{7}{2}$, 
\begin{equation} 
K^F_{7/2}(\rho_0,x_0;x) = K_{4}(\rho_0,x_{0};x) {1 \over 
\sqrt{\rho_0}}
\left( \rho_0 \gamma_{\hat5} +  {(x_{0}-x)}^{\mu}
	\gamma_{\hat\mu}	\right)
	\label{eq:prop-7/2} ,
\end{equation}   
which, suppressing all spinor indices, leads to  
\be
 \Lambda_{J}(x_0,\rho_0) = \int d^4 x K^F_{7/2}(\rho_0,x_{0};x) 
J_{\Lambda}(x) \, ,
\label{boundarylambda}
\ee
where $J_{\Lambda}(x)$ is a left-handed  boundary value of   
$\Lambda$ and acts as the 
source for  the composite operator  
${\widehat \Lambda}$  in the boundary $\calN=4$ SYM theory.
As a result, the classical action for the operator ${(\Lambda)}^{16}$
in the   
AdS$_{5}\times S^{5}$ supergravity action is
\begin{eqnarray}  
  S_{\Lambda} [J] &=& e^{-2\pi ({1\over g_s}  +  
  i  \chi_{_{0}})} g_s^{-12} \, 
  V_{S^5} \, \int \frac{d^{4}x_{0}d\rho_0}{\rho_0^{5}} \nonumber\\
  && t_{16} \prod_{p=1}^{16}  
  \left[ K_{4}(\rho_0,x_{0};x_{p}) \frac{1}{\sqrt{\rho_0}}
  \left( \rho_0 \gamma^{\hat 5} +  
  {(x_{0}-x_{p})}^{\mu} \gamma_{\hat\mu} \right) \,
  J_{\Lambda}(x_{p}) \right] \, ,
\label{source-l16} 
\end{eqnarray} 
where we have set $e^\phi = g_s$ and $ \chi = \chi_{_{0}}$ 
(since the scalar fields are taken to be 
constant in the \AdS5s5\ background) and $V_{S^5}=\pi^3$ 
is the $S^5$ volume. The 16-index invariant tensor $t_{16}$ is the same
as the one defined after (\ref{l16-sym}). 
The overall power of the coupling constant comes from the factor of 
$g_s^4$ in the measure (\ref{dmeasdef}) and the factor of
 $g_s^{-16}$ from the leading term in (\ref{fnexp}).
Using the dictionary (\ref{dict}) and 
 differentiating with respect to the chiral sources  
this result agrees with the  
expression (\ref{l16-sym}) obtained in the Yang Mills  
calculation, including the power of $g_{_{YM}}$. 

The remarkable agreement of the one D-instanton contribution
to the $\Lambda^{16}$ amplitude in type IIB superstring on $AdS_5\times S^5$
with the corresponding one-instanton contribution to
the sixteen-current correlation function in SYM theory, found in \cite{bgkr}
in the case of an $SU(2)$ gauge group, has been beautifully extended to any 
$SU(N)$ \cite{dm} and to $K$-instanton contributions \cite{mattis2}.
In the last paper, in particular, it is neatly clarified how the 
relevant YM-instanton moduli space reduces to a $K$-fold symmetric product
of $AdS_5\times S^5$ in the large $N$ limit and then collapses to a single 
copy of $AdS_5\times S^5$ due to an effective attraction among the various
instantons. This is precisely what was expected for the moduli
space of type IIB D-instantons on $AdS_5\times S^5$!
A non-renormalization theorem may protect the 16-fermion amplitude
and all other type IIB superstring amplitudes related to this 
by supersymmetry from $\ap$ corrections so that 
they should agree with their SYM counterparts for finite $N$ as well. 
Supersymmetry arguments and the precise agreement for the 16-fermion 
correlator in SYM with the correspondent 16-fermion
amplitude in type IIB superstring guarantee the agreement for certain
four-point correlators whose correspondence with type IIB amplitudes is less 
straightforward to see \cite{bgkr}. We will not pursue this line
of investigation any further if not for the purposes of showing the peculiar 
logarithmic behavior of some four-point functions of scalar composite 
operators. Before doing that, we will review to what extent the  
information about the charge-one D-instanton term can be directly 
extracted from a semi-classical type 
IIB supergravity computation around a D-instanton background. 

\section{The D-instanton solution in \AdS5s5}

In flat ten-dimensional Euclidean space the charge-$K$  
D-instanton solution is a finite-action Euclidean 
supersymmetric (BPS--saturated) solution  in which
the  metric is flat in the Einstein frame
but the complex scalar $\tau = \chi + i 
e^{-\phi}$ has a nontrivial profile with a singularity at the 
position of the D-instanton.   The (Euclidean) 
R--R scalar is related to the dilaton  by the BPS condition  
$\partial_\Sigma \chi = \pm i 
\partial_\Sigma e^{-\phi}$. The solution for the dilaton profile, 
found in \cite{ggp} and corrected in \cite{bgkr}, is
\be 
e^{\hat \phi^{(10)}}  = g_s + { 3 K {\al}^4 \over \pi^4 |X-X_0|^8} \;  
\label{tendim} 
\ee
and satisfies the ten-dimensional 
Laplace equation, $\partial^2e^\phi =0$,  outside  an
infinitesimal sphere centered on the point $X_0^\Lambda$ (where 
 $X^\Lambda$ is the ten-dimensional coordinate 
and $X_0^\Lambda$ is the location of the D-instanton).  
In (\ref{tendim}) $g_s$ is the asymptotic value of the string coupling 
and the second term has a quantized coefficient in virtue of 
the Dirac--Nepomechie--Teitelboim condition 
that quantizes the charge of an electrically charged   
$p$-brane and of its magnetically charged $p'$-brane dual 
\cite{nepoteit}. 
In solving the equations of motion of the type IIB 
theory for a D-instanton in Euclidean  \AdS5s5,  
the BPS condition requires 
\be 
e^{-\phi} g^{\Lambda\Sigma} \nabla_\Lambda \nabla_\Sigma e^\phi = 0 \quad .
\label{disteq} 
\ee 
In the Einstein frame, the Einstein equations are unaltered by
the presence of the D-instanton (because the associated Euclidean stress 
energy tensor vanishes) so  that the metric of \AdS5s5\ remains a solution.
The D-instanton equation (\ref{disteq}) is easy to solve
using the conformal flatness of the \AdS5s5\ metric (\ref{cfmetric}).
In Cartesian coordinates\footnote{The $y^i$ here are related to the $y^i$ in
(\ref{d3brane}) by an inversion.} 
$X^\Lambda=(x^\mu,y^i)$,  (\ref{cfmetric}) reads 
$ds^2 = L^2\rho^{-2} dX\cdot dX$, where $\rho^2 = y \cdot y$ and the
solution of (\ref{disteq}) with a constant asymptotic behavior is of the form
\be
\label{conflat}
e^{\hat \phi}  = g_s +  { \rho_0^4 \rho^4\over L^8} 
\left(e^{\hat \phi^{(10)}} - g_s \right),
\ee
where  $\rho_{_0}^2 = y_{_0} \cdot y_{_0}$ and $e^{\hat \phi^{(10)}}$ 
is the same harmonic function that appears in the flat ten-dimensional 
case, (\ref{tendim}). In evaluating D-instanton dominated  amplitudes 
we are interested in the case in which the point
$X = (x,y)$ approaches the boundary ($\rho \to 0$).
It is then necessary to rescale the 
dilaton profile (just as it is necessary to rescale the scalar
bulk-to-bulk propagator, \cite{gkp,wittone}) so that the combination
\be
\rho^{-4} \left(e^{\hat \phi} - g_s \right) =   
{3 K(\alpha')^4 \over L^8 \pi^4}{  \rho_0^4\over  ((x-x_0)^2 + 
\rho_0^2)^4 }  \, , 
\label{classd}
\ee
will be of relevance in this limit.     

As mentioned earlier, the correspondence with the YM instanton 
follows from the fact that $  \rho_0^4/ ((x-x_0)^2 + 
\rho_0^2)^4  = K_4$
 is proportional to the instanton number  density, $(F_{(0)}^-)^2$, 
in the $\calN=4$
 SYM theory.  Strikingly, the  scale size of the YM 
 instanton is replaced by the distance $\rho_0$ of the D-instanton 
from the boundary. This is another indication of how the geometry of the
 SYM theory is encoded in the type IIB superstring.   Note,
 in particular,  that 
as the D-instanton approaches the boundary $\rho_0 \to 0$, 
the expression for  $\rho^{-4} e^{\hat \phi}$ 
reduces to a $\delta$-function that corresponds to a zero-size YM instanton.  

Using the BPS condition, 
the  action for a D-instanton of charge $K$ can be written as 
\be
\label{actint}
S_K = {L^{10}\over (\alpha')^4} \int  
{d\rho d^4x d^5 \omega \over \rho^5}  g^{\Lambda\Sigma} 
\left( \partial_\Lambda \hat\phi \right) 
\left( \partial_\Sigma \hat\phi \right) \: ,
\ee
which reduces to an integral over the boundary of \AdS5s5\ and the surface 
of an infinitesimal sphere centered on the D-instanton at $x=x_0$, $y=y_0$.  
Substituting for $\hat\phi$ from (\ref{conflat}) gives 
\be 
S_K = {2\pi |K|\over g_s} \, ,
\label{dinstact} 
\ee 
which is the same answer as in the flat ten-dimensional case.  
Remarkably, the boundary integrand is {\it identical} 
to the action density  of the standard four-dimensional YM instanton. 

The D-instanton contribution to the amplitude with
sixteen external dilatinos  may then be obtained directly by
semi-classical quantization around the classical
D-instanton solution in \AdS5s5. 
The leading instanton contribution can be determined by applying
supersymmetry transformations to the classical scalar field  
background (\ref{classd}).   Since 
the D-instanton background breaks half the supersymmetries the
relevant transformations are those in which the supersymmetry
parameter corresponds to the  Killing  
spinors for the sixteen broken supersymmetries.  
 
Up to the overall constant factor, the  amplitude computed in \cite{bgkr}
agrees with (\ref{source-l16}) and therefore with (\ref{l16-sym}).   
In similar manner the instanton profiles of all the fields in the
supergravity multiplet follow by applying the broken supersymmetries
to the D-instanton solution just as they do in the flat
ten-dimensional case \cite{greengut}.  The one D-instanton
contributions to any correlation function can then be determined 
in principle.

\section{Logarithmic behavior of four-point functions}

In this last part of the talk we would like to show some peculiar behaviors 
of the exact non-perturbative correlators of superconformal currents. 
To this aim we shall consider four-point functions.
The prototype correlation function of four superconformal 
currents is the correlation function  
of four stress tensors. However, due to its complicated 
tensorial structure even the free-field expression for this correlator
is awkward to compute and it is much simpler to consider 
correlations of four gauge-invariant  composite scalar operators, 
such as the ${\bf 20}$ scalars ${\cal Q}^{ij}$ defined in (\ref{qdef}) 
or the scalar $SU(4)$ singlet ${\cal C}(x)=\Tr(F^{-})^{2}$. 
After calculating correlation functions  of these 
currents  one can derive those of any other currents in 
the ${\cal N}$=4 supercurrent multiplet by making use of the  
superconformal symmetry \cite{howest}. 
Instead of considering the most general correlation function,
\be 
 \langle {\cal Q}^{i_1j_1}(x_1)  {\cal Q}^{i_2j_2}(x_2)  
 {\cal Q}^{i_3j_3}(x_3) {\cal Q}^{i_4j_4}(x_4)\rangle \quad ,
\label{qfourpoint} 
\ee
in order to evaluate one-instanton contributions 
in the most convenient fashion, it is useful to exploit the
$SU(4)$ symmetry and only compute the correlation function 
of two  $\varphi^2$ and two ${\overline{\varphi}}^2$. 
It will then prove useful to work in the manifest
${\cal N}$=2 supersymmetric description of (\ref{entwo}). 
After performing the necessary Wick contractions, 
the free-field expression for this particular correlator is 
\begin{equation}
   \langle\varphi^2(x_1){\varphi}^2(x_2){\overline {\varphi}}^2(x_3) 
   {\overline{\varphi}}^2(x_4)
   \rangle_{_{\mbox{\tiny{free}}}} = \frac{1}{(4 \pi^2)^{4}} 
   \left({4 N^{2} \over x_{13}^4 x_{24}^4} 
   + {4 N^{2} \over x_{14}^4 x_{23}^4} +  
   {16 N \over x_{41}^2 x_{13}^2 x_{32}^2  x_{24}^2} \right) \, .
   \label{corrpert}
\end{equation}
 
The computation of the one-instanton contribution to this Green 
function is straightforward using standard techniques \cite{akmrv}. 
The semi-classical approximation obtains after replacing the 
fields $\varphi$ with the expression 
\be 
\varphi(x) \rightarrow \varphi_{(0)}(x) =  \frac{1}{2\sqrt{2}}
\varepsilon_{uv} \zeta^{u}_{+} \s^{\mu\nu} \zeta^{v}_{+}   
F^-_{(0)\mu\nu}  \: , 
\label{scalarsol} 
\ee
and a similar one for $\overline{\varphi}$. These are 
the leading nonvanishing terms that     
result from  Wick contractions with Yukawa-coupling terms lowered 
from the exponential of the action until a sufficient number of 
fermion fields are present to saturate the fermionic integrals.    
Of course, these expressions can also be obtained 
directly from the supersymmetry transformations (\ref{transs}) 
by acting twice on $F^{-}_{(0)\mu\nu}$ with the broken supersymmetry 
generators. 
After some elementary Fierz rearrangements on the fermionic  
collective coordinates the fermionic integrations can be performed 
in the standard manner to yield
\be  
G_{4}(x_{p}) = g_{_{YM}}^8
\,e^{-{8\pi^2 \over g_{_{YM}}^2} + i\theta_{_{YM}}} 
\,   \int {d\rho_0 d^4 x_{0} \over \rho_0^5} \: 
x_{12}^4 x_{34}^4 \, \prod_{p=1}^4 {\left( {\rho_0 \over 
{\rho_0^2 + (x_p - x_{0})^2}}  \right)}^4 . 
\label{collectint} 
\ee
As remarked earlier, the fact that  the  instanton form factor,
$\rho_0^4/[(\rho_0^2 + (x_p - x_0)^2]^4$, that enters this expression 
is identical to $ K_4$ in (\ref{greenfun}) is of the outmost significance in 
establishing of the AdS/Yang--Mills correspondence.
The integration  in (\ref{collectint})
resembles that of a standard Feynman diagram with momenta 
replaced by position differences and can be performed by introducing 
the appropriate Feynman parametrization. 
Up to an overall numerical constant, the five-dimensional integral 
then yields 
\be
G_{4}(x_p) =
g^{8}_{_{YM}}e^{-{8\pi^2 \over g_{_{YM}}^2} + i\theta_{_{YM}}}   
\int \prod_p \alpha_p^3 d\alpha_p \, 
\delta \left(1-{\sum}_q  \alpha_q \right) 
{x_{12}^4 x_{34}^4 \over 
(\sum_p \alpha_p \alpha_q x_{pq}^2 )^{8}} \: . 
\label{integral} 
\ee   
This integral can be obtained by acting with six derivatives on the 
`box-integral' with four massless external particles. Indeed 
\be  
G_{4}(x_p) =
g^{8}_{_{YM}} e^{-{8\pi^2 \over g_{_{YM}}^2} + i\theta_{_{YM}}}  \,
x_{12}^4 x_{34}^4 \prod_{p<q} 
{\partial \over \partial x_{pq}^2} B(x_{pq}) \, , 
\label{intbox} 
\ee   
where the box integral $B(x_{pq})$ turns out to be a combination of 
logarithms and dilogarithms \cite{bern}
\ba  
B(x_{pq}) &=& {1\over \sqrt{\Delta}} \left[ 
- {1\over 2} \log\left ({u_+u_-\over(1-u_+)^2(1-u_-)^2}\right) 
\log\left({u_+\over u_-} \right) \right.  \nonumber  \\  
&& \hspace{-1.8cm} - \left. \mbox{ Li}_2(1-u_+) + 
\mbox{Li}_2(1-u_-) -\mbox{ Li}_2\left( 1-{1\over u_-} \right) + 
\mbox{ Li}_2\left( 1-{1\over u_+} \right) \right] \; ,
\label{boxeplicit} 
\ea   
where   
\be 
\Delta = \det_{4\times 4}((x^2_{pq})) =    
X^2 + Y^2 + Z^2 - 2 XY - 2 YZ - 2 ZX 
\label{det} 
\ee   
and 
\be 
u_{\pm} = { Y + X - Z \pm \sqrt{\Delta} \over 2 Y} \; , 
\label{ratio} 
\ee   
with $X = x_{12}^2 x_{34}^2$, $Y = x_{13}^2 x_{24}^2$ and  
$Z=x_{14}^2 x_{23}^2$. 
 
Up to the overall factor $x_{12}^4 x_{34}^4$ needed for the correct
scaling, $G_{4}$ turns out to be a function of the two
independent superconformally invariant cross-ratios  $X/Z$ and $Y/Z$.
Although not immediately apparent, the expression for $B(x_{pq})$
is symmetric under any permutation of the external legs.

Unlike  correlation functions of elementary 
fields that are  infra-red problematic and gauge-dependent \cite{kov}, 
the above correlator is well defined at non-coincident points.  
Up to the derivatives acting on the box integral, the result is 
exactly the one expected for the correlator of four scalar operators  
of dimension  $\Delta = 2$ each. As observed in \cite{bgkr} it is not 
completely straightforward to compare the
detailed expression for this correlator with the one D-instanton contribution 
to  the ${\cal R}^4$ term in the type 
IIB  effective action around the AdS background.
In this respect, it is easier and perhaps more interesting from a 
`phenomenological' viewpoint to consider the correlator of four scalar 
`glueball' fields ${\cal G}=\mbox{Re}\,{\cal C}=\Tr (F^2)$. 
To leading order and up to an overall numerical constant, 
the one-instanton contribution to
this four-point correlation function is expected to be
\be
\la {\cal G}(x_1){\cal G}(x_2){\cal G}(x_3){\cal G}(x_4) \ra_{_{K=1}} =
g_{_{YM}}^8 \, e^{-{8\pi^2 \over g_{_{YM}}^2} + i\theta_{_{YM}}} 
\,   \int {d\rho_0 d^4 x_{0} \over \rho_0^5} \: 
\prod_{p=1}^4 {\left( {\rho_0 \over 
{\rho_0^2 + (x_p - x_{0})^2}}  \right)}^4 . 
\label{fourc}
\ee
and coincides with (\ref{collectint}) 
after removing the factor $x_{12}^4 x_{34}^4$,
consistently with superconformal invariance. Differently from (\ref{l16}),
the correlators (\ref{fourc})
and (\ref{collectint}) respect the $U(1)_B$ symmetry \cite{int} and 
receive an analogous (complex conjugate) contribution from an anti-instanton. 
As observed by other groups \cite{log},
this expression shows a logarithmic singularity in the limit 
$x_{ij}\to 0$ for each pair of external points. This may be either a 
drawback of the lowest-order approximation or an intrinsic pathology of the 
theory. The logarithms by themselves do not violate the superconformal 
symmetry of the theory that is manifest in the final expression of the 
four-point functions given in \cite{bgkr} and reviewed above. In some 
respect, logarithmic behaviors in four-point functions should not sound 
unexpected in a superconformal theory such as ${\calN}$=4 SYM theory that 
contains a large number of primary fields all of whose 
(protected) dimensions are integer. 
Rather they may signal a violation of the Operator Product
Expansion. The latter may be due to the presence of an infinite tower 
of stable BPS dyons that collapse to vanishing mass and size 
in the superconformal phase.  
From the Euclidean viewpoint this effect may be due to the presence of 
a gas of instantons of vanishing size.
This problem is under active investigation \cite{bkrs}.

\section{Discussion} 
 
The fact that YM-instanton and type IIB D-instanton effects 
appear to match so closely should be interpreted as a strong support to the  
conjectured correspondence put forward by Maldacena
\cite{maldacena}.  Most notably, we have seen that the YM 
instanton scale size has a natural interpretation as the position of 
the D-instanton in the extra dimension transverse to the 
four-dimensional boundary space-time. This reflects the fact that the 
one-instanton moduli spaces in both cases contain an $AdS_5$ factor.
The  additional $S^5$-directions, which are not at all apparent in standard
${\cal N}$=4 SYM perturbation theory, are associated to auxiliary bosonic
`collective coordinates' \cite{mattis2}.

The remarkable agreement that has been found in \cite{bgkr}
for the lowest non-trivial value of $N$ ($N=2$) and its extension to any $N$ in
\cite{dm} neatly show the computational power of instanton calculus.
Indeed it would have been hopeless to try a matching of lower derivative terms 
in the type IIB effective action with the planar series to corresponding SYM
correlation functions.
On the contrary, the agreement between non-perturbative
and thus highly suppressed contributions to some protected interactions
may even lead one to suggest that the original conjecture be valid for
finite, indeed any, $N \geq 2$. For consistency one should expect
a truncation to $\ell\leq N$ of the KK spectrum in the type IIB superstring
compactification on $S^5$ for finite radius. Worldsheet unitarity 
requirements analogous to the ones familiar in $SU(2)$ WZW models may underlie
these stringy effects. Indeed this seems to be the case in the correspondence
between two-dimensional SCFT's and type IIB superstring on $AdS_3\times S^3$
\cite{msetal}. However, despite some interesting work \cite{gsads} the 
quantization of the GS-type action governing the dynamics of the type IIB 
superstring around $AdS_5\times S^5$ does not seem to be at reach in the 
near future.

For this reason, the evaluation of $K$-instanton contributions in 
${\cal N}=4$ SYM theory in the large $N$ limit and its agreement with type 
IIB D-instanton effects \cite{mattis2} may be considered as the strongest 
support to the correspondence found so far.
A detailed analysis of the contribution of multiply-charged ($K>1$) instanton 
configurations for finite $N$ should clarify what survives of the
correspondence for finite $N$. The quantitative agreement of the
charge-$K$ D-instanton coefficients in the type IIB modular functions
$f_n$ and the corresponding coefficients in SYM theory for large $N$
is a promising step in this direction \cite{mattis2}. 

An alternative approach to the non-perturbative computations performed so far 
\cite{bgkr,dm,mattis2} consists in viewing the instanton-dominated correlators 
as a topological subsector of the ${\cal N}$=4 SYM theory.  
In this respect, it is worth noticing that the  
modular-covariant non-holomorphic functions of the complexified 
couplings, $f_n(\tau,\bar\tau)$, 
that appear in the type IIB D-instanton description are amazingly
similar to some expressions found in the computation of the Witten index for 
topologically twisted ${\cal N}$=4 theories on curved manifolds 
\cite{vafawitten}.

\vspace{1cm}
 
{\bf Acknowledgments}

\vspace{0.3cm}
 
\noindent
The content of this talk is almost entirely based on work done  
with Michael Green and Giancarlo Rossi, that we thank for
a very enjoyable collaboration.
We acknowledge useful e-mail correspondence with Michael Mattis and 
interesting discussions with Yassen Stanev and 
the partecipants to the RFBR-INTAS School on
{\it Advances in Quantum Field Theory, Statistical Mechanics and Dynamical
Systems} held in Como, to the Summer Institute {\it Miramare '98} held 
in Trieste and to the workshop {\it Conformal Field Theory of D-branes} 
held in Hamburg, where various versions of this talk have been presented.

\end{document}